\documentclass[11pt,twoside]{article}
\usepackage{asp2010}
\usepackage{graphicx}
\usepackage{natbib}

\resetcounters

\begin{document}

\title{PlotXY: a high quality plotting system for the Herschel Interactive
Processing Environment (HIPE), and the astronomical community}
\author{Pasquale Panuzzo$^1$, Jinjing Li$^{2,3}$, and Emmanuel Caux$^{4,5}$
\affil{$^1$CEA Saclay, Laboratoire AIM, Irfu/SAp, Orme des Merisiers, F-91191 Gif-sur-Yvette, France}
\affil{$^2$National Astronomical Observatories, CAS, 20A Datun Road, Chaoyang District, 100012 Beijing, China}
\affil{$^2$GSegment Space Technologies, Inc., Room 1513, CRD YinZuo, Wanda Plaza, ShiJIngShan District, 100040 Beijing, China}
\affil{$^4$Universit\'e de Toulouse, UPS-OMP, IRAP, Toulouse, France}
\affil{$^5$CNRS, IRAP, 9 Av. colonel Roche, F-31028 Toulouse cedex 4, France}
}

\begin{abstract}
The Herschel Interactive Processing Environment (HIPE) was developed by the European Space Agency (ESA)
in collaboration with NASA and the Herschel Instrument
Control Centres to provide the astronomical community a complete environment to process and analyze
the data gathered by the Herschel Space Observatory.
One of the most important components of HIPE is the plotting system (named PlotXY) that we present here.
With PlotXY it is possible to produce easily high quality publication ready 2D
plots. It provides a long list of features, with fully configurable components, and interactive zooming.
The entire code of HIPE is written in Java and is open source released under the GNU
Lesser General Public License version 3. A new version of PlotXY is being developed to be independent
from the HIPE code base; it is available to the software development community
for the inclusion in other projects at the URL \url{http://code.google.com/p/jplot2d/}.
\end{abstract}

\section{Introduction}
The Herschel Space Observatory\footnote{{\it Herschel} is an ESA space observatory with science 
instruments provided by European-led Principal Investigator consortia and with important participation from NASA.} 
\citep{2010A&A...518L...1P} is an ESA cornerstone mission 
developed in collaboration with NASA. It consists of a space 
telescope with a primary mirror of 3.5 meters equipped with three 
scientific instruments \citep[HIFI, PACS, and SPIRE;][]{2010A&A...518L...6D,2010A&A...518L...2P,2010A&A...518L...3G} to observe the 
universe in the far infrared and submillimeter spectral region 
(from 50 to 670$\mu$m). 

Herschel was launched on 14th May 2009 together with the ESA 
Planck microwave observatory; its mission lifetime is expected to 
be of 3.5 years. Herschel is operated as an observatory open to the 
international scientific community, with 1/3 of the total time 
reserved to the consortia institutes that built the scientific 
instruments. More information on scientific goals and results of 
the Herschel mission can be found on the Herschel Science Center 
website\footnote{\url{http://herschel.esac.esa.int/}}. 

The development, management and the scientific exploitation of 
a complex mission as Herschel required the development of a 
large software system (the Herschel Common Software System, 
HCSS). The HCSS was developed by a large pool of developers 
and scientists of ESA, NASA and Instrument Control Centres 
institutes, distributed in many countries (Europe, USA, Canada, 
China). The HCSS is written in Java and it is open source, 
released under the GNU Lesser General Public License v3. 

The Herschel Interactive Processing Environment \citep[HIPE,][]{2010ASPC..434..139O}
is a major component of the HCSS that was 
developed to provide the astronomical community and the 
instrument teams a complete environment to process and analyze 
the data gathered by Herschel. 

As all data analysis systems, HIPE needs a tool to produce plots. 
Several open source java plotting libraries are already available 
(e.g. JFreeChart and the SGT Library), however the astronomical 
community has its own standards for the style of plots, e.g. the 
plot has to be in a closed box, with axis ticks oriented inward, etc. 
as in plots produced with IDL,\footnote{\url{http://www.ittvis.com}} widely used by 
the astronomical community. 

In order to meet these requirements, after the evaluation of 
available libraries, we implemented our own plotting system. The 
development started from the SGT library code, which was later 
fully replaced with new code. 

This new plotting system, called PlotXY, was developed by 
Jingjing Li as one of the Chinese contributions to the mission. P. 
Panuzzo and J. Bakker contributed to the code development. This 
poster presents PlotXY features, architecture and future 
developments.


\begin{figure}[!ht]
\centerline{\includegraphics[width = 4.3cm]{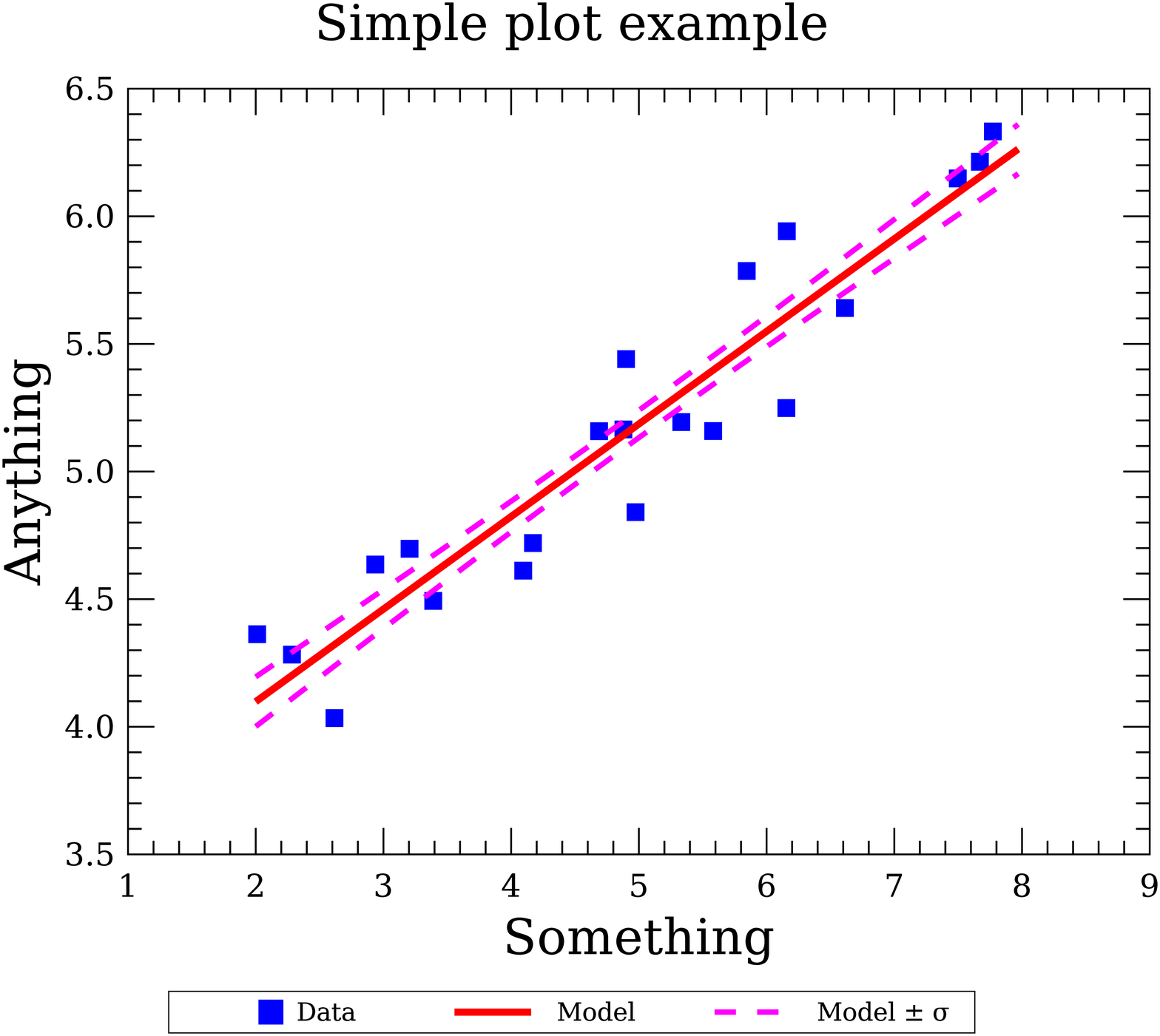}\includegraphics[width = 4.3cm]{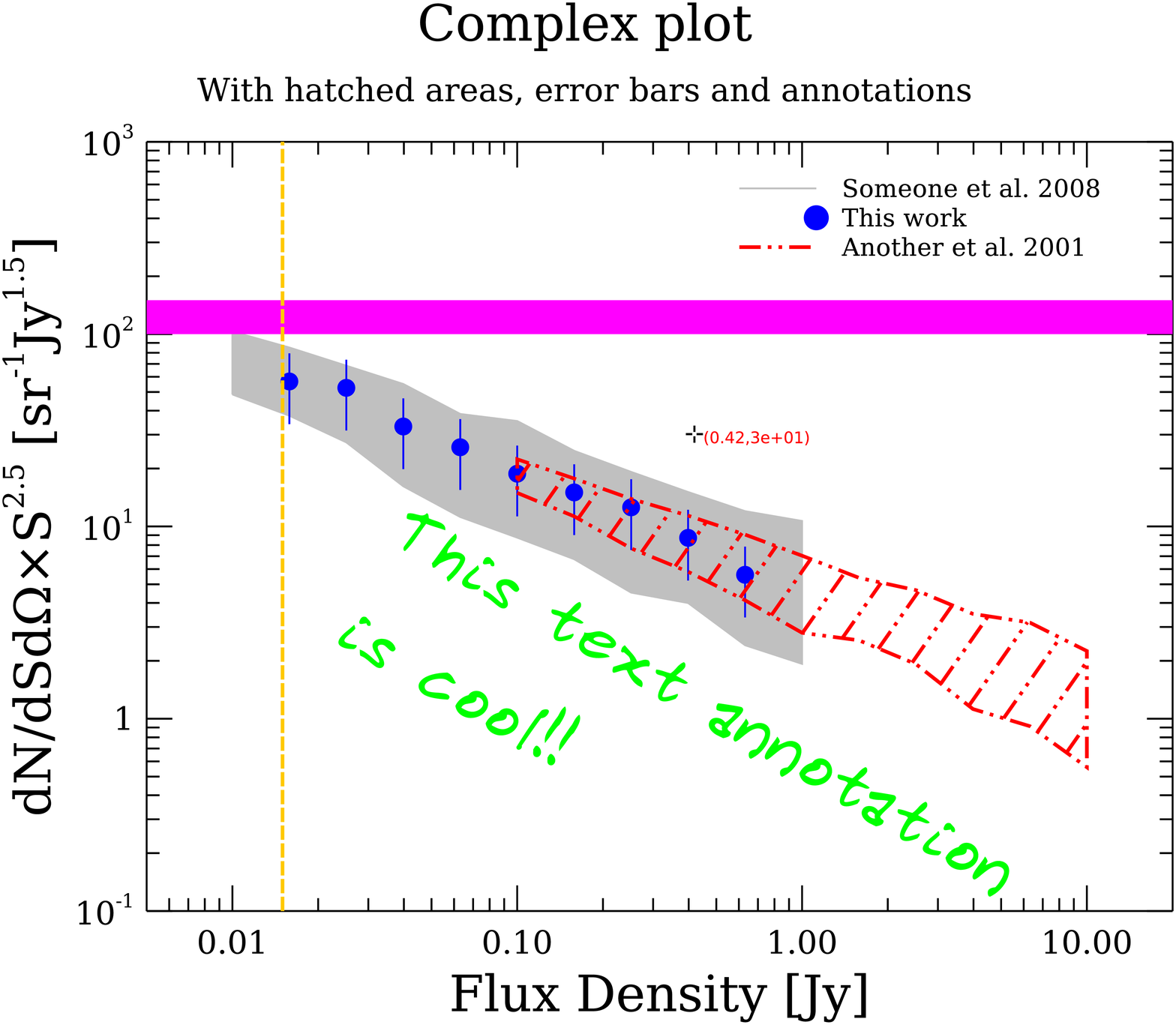}\includegraphics[width = 4.3cm]{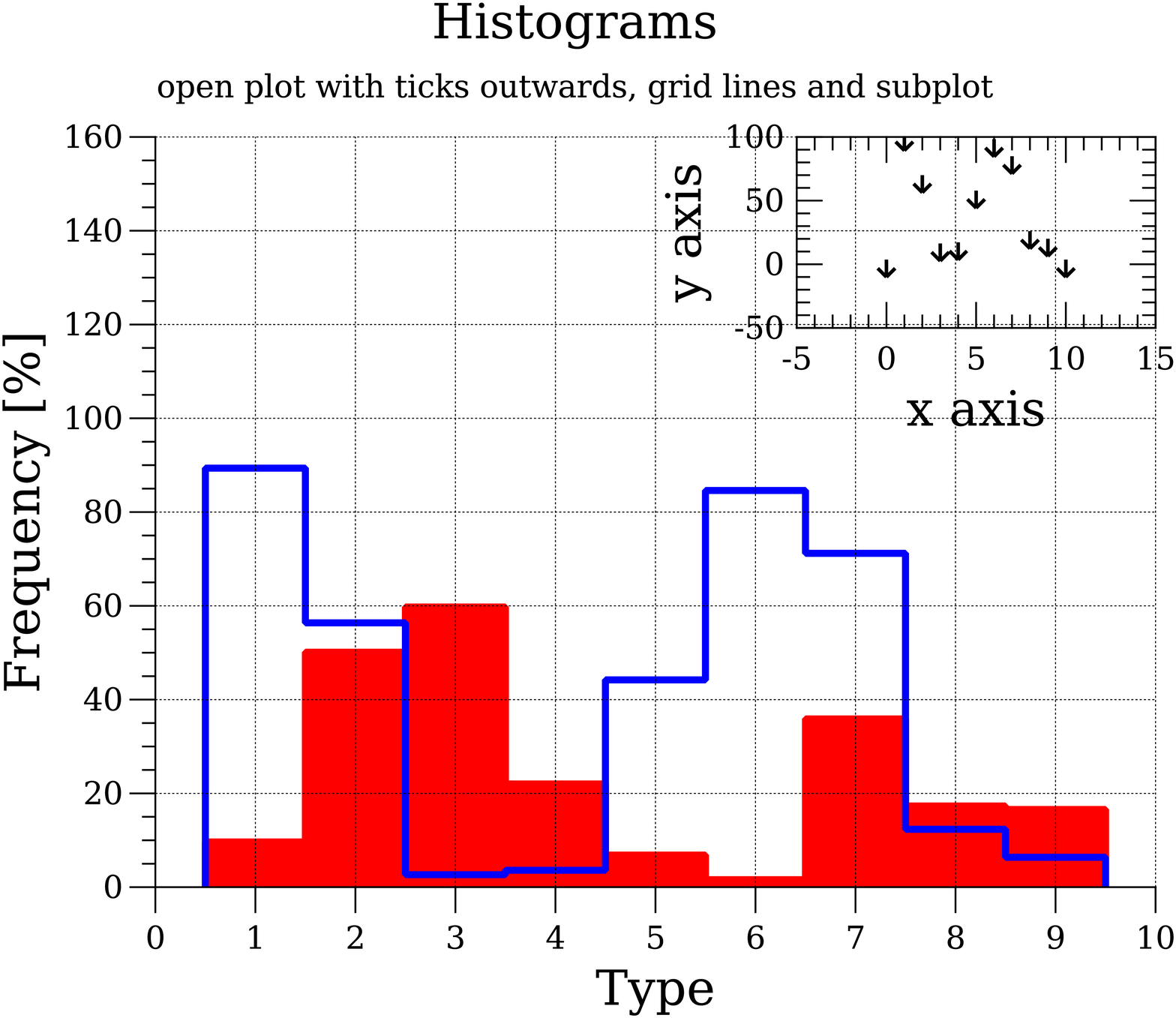}}
\centerline{\includegraphics[width = 7.0cm]{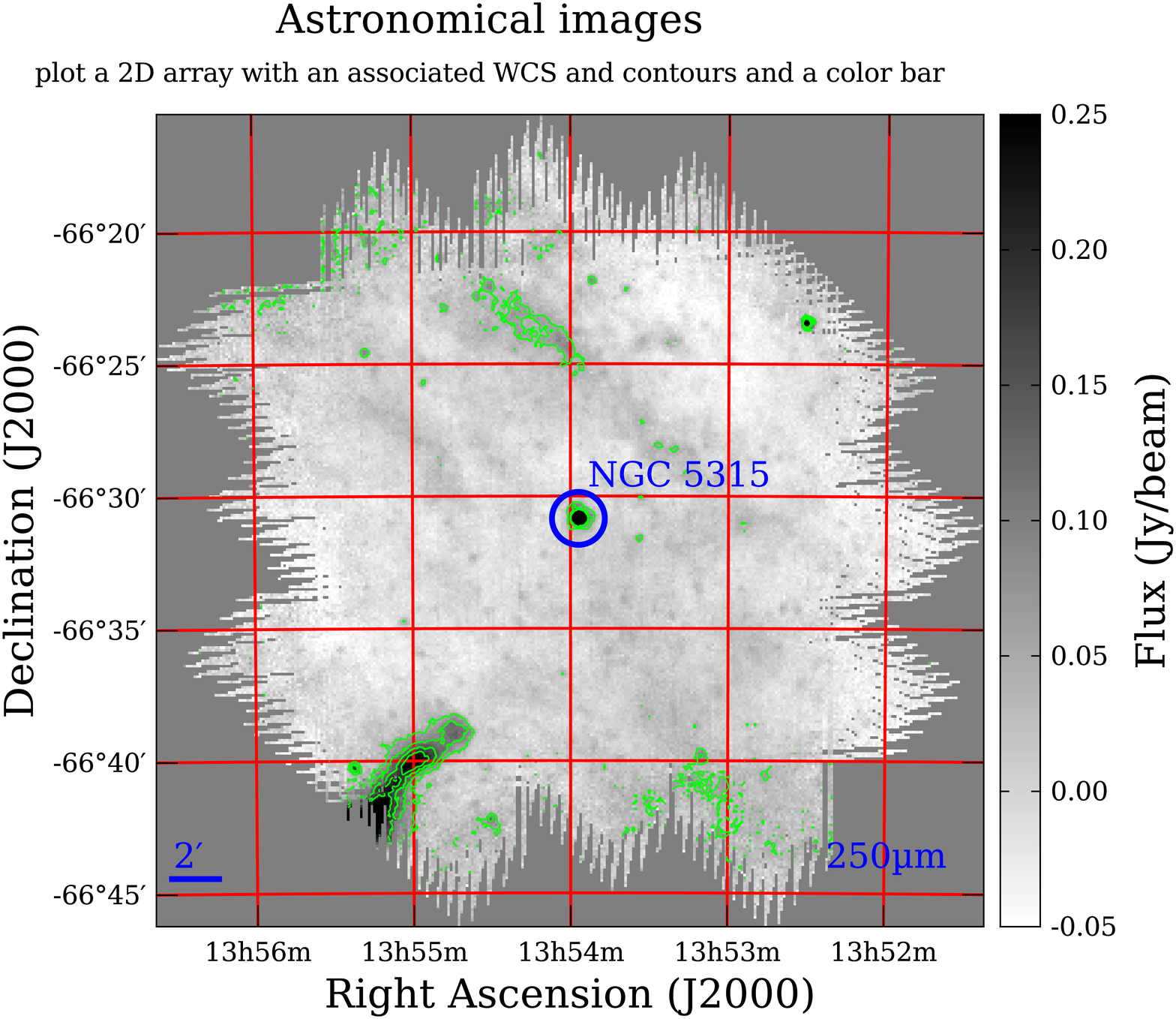}\includegraphics[width = 4.5cm]{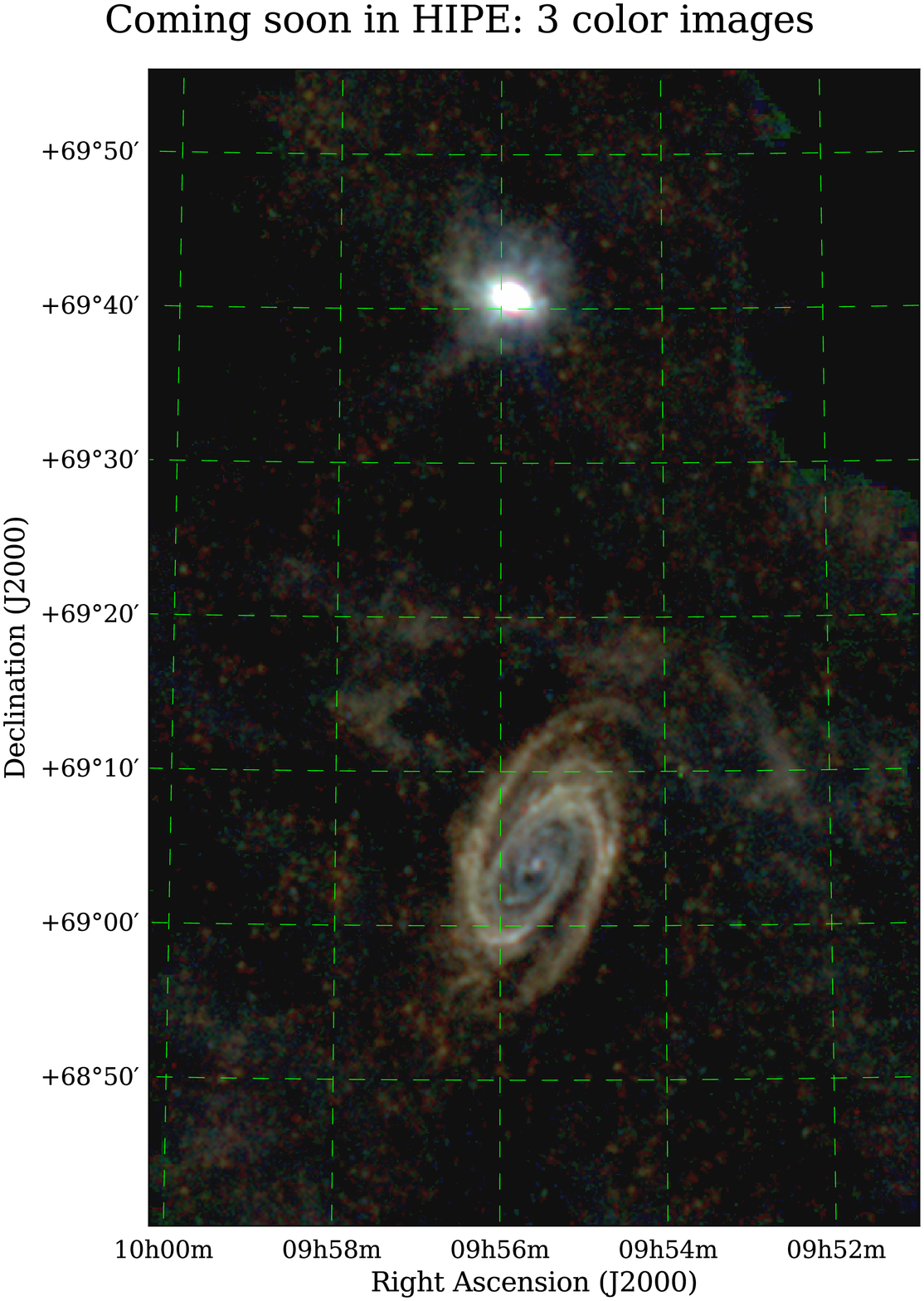}}
\caption{Various examples of plots generated with PlotXY.}
\label{fig_plots}
\end{figure}

\section{PlotXY Features}

PlotXY allows to plot data on a 2D Cartesian coordinate
system. It doesn't support 3D plots, pie plots, nor polar plots.
The type of data that can be displayed are: a) pairs of 1
dimensional arrays (X,Y), b) pairs of 1 dimensional arrays with
errors, c) 2D scalar arrays (like an intensity map, see Fig. \ref{fig_plots} left-bottom panel), d)
3 colors images (to be included in HIPE soon, Fig. \ref{fig_plots} right-bottom panel). It could
be easily extended to display vectorial fields.

Data are displayed in a drawing box delimited by axes (see
 Fig. \ref{fig_plots} left-top panel, for a simple example), as is typically required in
astronomical publications. It is however possible to make open
plots by removing some of the axes (e.g. Fig.~\ref{fig_plots} right-top panel). Axes can be of
different types: linear, logarithmic, astronomical coordinates,
date. It is possible to define different scales in each axis of the
drawing box, set the number or the position of minor and major
ticks, as well the labels.

Each set of data is displayed in a layer. One or more layers are
 associated to a drawing box, i.e. a subplot. One or more subplots
are associated to a plot (see Fig.~\ref{fig_plots} right-top panel for an example of a plot with
2 subplots).

Each layer has an associated style that describes how the data
 shall be displayed, i.e. the chart type (histogram or normal),
symbols, line type (solid, marked, symbols, dashed), color,
stroke etc.
It is also possible to add several type of annotations to a plot,
 like a text, an horizontal or vertical line, a rectangle area 
(Fig.~\ref{fig_plots}, center-top). Finally it is also possible to add grid lines.
All properties of all components (e.g. colors, fonts, axis ranges,
 line styles etc.) can be changed interactively via a property panel
or via methods calls.

The system provides an interactive zoom, i.e. the user can
 select with the mouse a region of interest and zoom in it, or
zoom in and out with the mouse wheel.
The user can also obtain the coordinates of a mouse click on
 the plot.
PlotXY also provides a batch mode to postpone rendering
 when many plotting properties and data are changed at the same
time.
Plots can be exported to JPEG, PNG, EPS and PDF files.
Plots can be shown in a dedicated window or used inside a
  container of an external GUI.

\section{Architecture}
\begin{figure}[!ht]
\centerline{\includegraphics[width = 10.8cm]{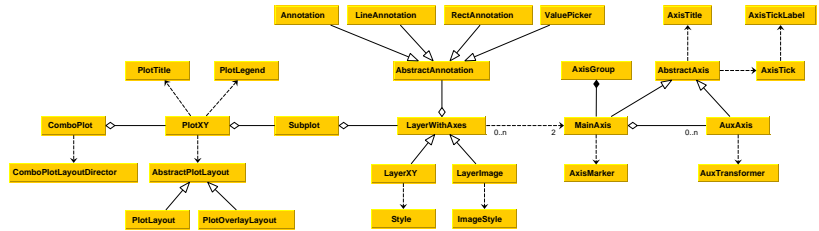}}
\caption{UML class diagram of PlotXY user classes.}
\label{fig_uml}
\end{figure}

The implementation is divided in 2 parts: 1) the classes used by the user to create the plots and 2) the
engine classes that take care of the rendering of the data. Fig. \ref{fig_uml} shows an UML class diagram of user
classes.

The separation of the user classes from engine classes allows for a greater flexibility and modular
development. In this way the engine classes can be modified without affecting the user API and, in
principle, the design allows to have several sets of switchable engines.

User actions (like changing a line style) trigger actions of the engines (like new rendering of the data).
These commands are executed in a thread-safe mode by executing them in the EDT (Event Dispatcher
Thread).
In order to get good performances when the user changes continuously the data to be displayed, we use
multi-thread rendering, i.e. each new request of rendering due to a change of the data is executed in a
new thread.

\section{The jplot2d project}

Jplot2d is a re-implementation of the Herschel PlotXY. It is intended to be a
plotting library that can be used by any other java application and it is
independent from the rest of Herschel code base.

Jplot2d has an improved design and more features, like:
\begin{itemize}
\item Thread-safe by synchronization in proxy
\item Rendered result of plot components can be cached
\item Display-dependent components separated from core classes, so that it can
be used in servlets
\item Neat architecture with less classes and more flexibility. 
Plots can contain other plots (no need of specific classes for subplots), layers can contains multiple sets of data
\end{itemize}
Jplot2d is still in early development. It is released under the LGPL3 and it
can be downloaded from URL \url{http://code.google.com/p/jplot2d/}.

\acknowledgements HCSS and HIPE are joint developments by the Herschel Science Ground Segment Consortium, consisting of ESA, the NASA Herschel Science Center, and the HIFI, PACS and SPIRE consortia.

\bibliography{P110}

\end{document}